\documentclass[prl,twocolumn,showpacs,superscriptaddress,preprintnumbers]{revtex4}

\usepackage{graphicx}
\usepackage{dcolumn}
\usepackage{bm}

\begin{document}

\preprint{}
\title{Drastic Reduction of Shot Noise in Semiconductor Superlattices}
\author{W. Song}
\affiliation{Department of Physics and Astronomy, State University of New York at Stony Brook, Stony Brook, NY, 11794-3800}
\affiliation{Korea Research Institute of Standards and Science, Daejeon 305-340, Korea}
\author{A.K.M. Newaz}
\affiliation{Department of Physics and Astronomy, State University of New York at Stony Brook, Stony Brook, NY, 11794-3800}
\author{J. K. Son}
\affiliation{Department of Physics and Astronomy, State University of New York at Stony Brook, Stony Brook, NY, 11794-3800}
\affiliation{Samsung Advanced Institute of Technology, Suwon 440-600, Korea}
\author{E.E. Mendez}
\email[e-mail address:\,\,\,]{emendez@notes.cc.sunysb.edu}
\affiliation{Department of Physics and Astronomy, State University of New York at Stony Brook, Stony Brook, NY, 11794-3800}
\date{Version: \today }
\date{\today}

\begin{abstract}

We have found experimentally that the shot noise of the tunneling current $I$ through an undoped semiconductor superlattice is reduced with respect to the Poissonian noise value $2eI$, and that the noise approaches 1/3 of that value in superlattices whose quantum wells are strongly coupled. On the other hand, when the coupling is weak or when a strong electric field is applied to the superlattice the noise becomes Poissonian. Although our results are qualitatively consistent with existing theories for one-dimensional mulitple barriers, the theories cannot account for the dependence of the noise on superlattice parameters that we have observed.

\end{abstract}

\maketitle

% body of paper here - Use proper section commands

The tunneling of electrons through a potential barrier, although itself a wave-mechanics phenomenon, also reveals the particle-like character of the charge. Thus the fluctuations of an average current, $I$, of uncorrelated electrons that tunnel through a single barrier are governed by Poisson statistics, whose out-of-equilibrium noise ({\it{shot noise}}) power spectrum $S$ is given by $S_P = 2eI$, as in a vacuum tube or in the space-charge region of a $p-n$ semiconductor junction. On the other hand, significant deviations from the Poissonian value occur when tunneling takes place through two or more barriers, as a result of correlations in the motion of the electrons. 

It has been found that the shot noise of a double-barrier resonant-tunneling diode is reduced or enhanced relative to $S_P$ depending on whether the diode is in the quasi-linear or in the negative-differential-conductance (NDC) region of its current-voltage $(I-V)$ characteristic - a non-Poissonian behavior that results from charge accumulation in the well \cite{Li90,Ian98,Kuz98,Blanter00}. In the quasi-linear region, the ratio $S / S_P$ (frequently called Fano factor, $F$) determined experimentally has typically ranged from 1/2 (if the tunneling probabilities through the individual barriers are the same) to 1 (when those two probabilities are very different from each other). Most calculations have accounted for that range, regardless of whether the analysis is semiclassical or quantum-mechanical and of whether the tunneling process is sequential or coherent\cite{Chen91}. There have been some calculations, however, that predict that in a fully coherent process the Fano factor can be significantly smaller than 1/2 (Ref. 6), a conclusion that seems to be substantiated by a few experimental results\cite{Brown92}. 

\begin{figure}[thp!]
  \centering
  \includegraphics[width=3.6 in,angle=0,clip=]{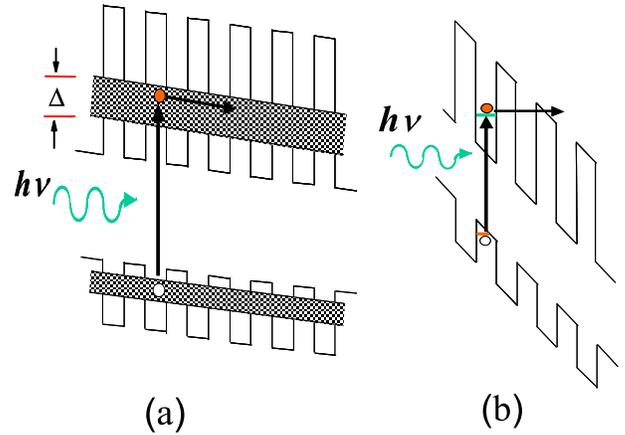}
  \caption[Potential Profile of a superlattice]{\label{fig:fig1} Schematic potential profile felt by a photo-generated electron tunneling through an undoped semiconductor superlattice of miniband width $\Delta$, in the low- (a) and high-electric (b) field limits, when the field is perpendicular to the superlattice's material interfaces. The electrons are pumped from the valence to the conduction band by photons whose energy lies between the bandgaps of the two semiconductors making the superlattice.}
\end{figure}

Theorists have generalized their calculations to multiple-barrier structures, using them to analyze the current fluctuations in mesoscopic metals. It has been observed that in a diffusive conductor shorter than its inelastic mean free path the Fano factor of the noise spectrum is $F = 1/3$ (Ref. 8), and calculations using either a full quantum-mechanical\cite{Been92} or a semiclassical\cite{Nag92} formalism have accounted for that value, with the implication that phase coherence is not essential for noise suppresion. In particular, a semiclassical model in which a metal is envisioned as an array of many potential barriers predicts that the larger the number, $N$, of tunneling barriers the larger the noise suppression; when all the barriers are identical, in the $N\rightarrow\infty$  limit the shot noise should approach 1/3 of the Poissonian value\cite{De95}.
 
Somewhat surprisingly, until now there has been no attempt to assess experimentally the limits of such a model using, for example, multi-barrier structures made out of two semiconductors, whose barrier height, width, and spacing can be controlled at will. For example, by varying the separation between barriers (or any of the other two parameters) it should be possible to probe the transition from coherent to sequential tunneling, and by applying an electric field the effective number of barriers that electrons tunnel can be varied in a controlled way.

We report here that the shot noise of a semiconductor superlattice, consisting of many (typically more than 20) alternating barriers and wells, approaches the $F = 1/3$ limit, provided the separation between barriers is small (1.5 nm) and the heterostructure is not driven very far from equilibrium, so that the barriers remain almost identical to each other (low-electric-field regime). On the other hand, in the high-field regime the shot noise is Poissonian, as expected when tunneling occurs through very different barriers. The shot noise observed in the transition region from one regime to another, however, goes beyond the simple predictions for a one-dimensional chain of barriers\cite{De95}.
 
Traditionally, electronic noise in tunneling transport has been studied by injecting electrons into an undoped (or intrinsic, $i$) semiconductor heterostructure via a voltage applied between $n$-type electrodes at the two ends of it. When the focus is on field effects, it is advantageous to work with a $p-i-n$ configuration in which the structure of interest is placed in the space-charge region of a $p-n$ diode, where large fields can be reached by applying a reverse bias to the diode without drawing any significant current. However, this creates a problem when trying to measue the current noise. We have circumvented this drawback by injecting carriers into the conduction band of the $i$ region via optical excitation of electrons from the valence band. In short, we have measured the shot noise of the photocurrent generated in a $p-i-n$ diode whose intrinsic region is formed by superlattice.

Electronic transport in a superlattice is a complex process, but for the purpose of this work it can be simplified into three regimes, in relation to a localization electric field $E_c$ defined by $E_c = \Delta/ eD$, where $\Delta$ is the width of the miniband of allowed energies and $D$ is the superlattice period\cite{Mendez88}. The two extreme regimes are sketched in Fig.1. At very low fields $(E \ll Ec)$, the superlattice miniband has not yet been destroyed by the electric field and optically injected carriers reach the $n$ electrode ("reverse" current) through miniband transport. At very high fields $(E \gg Ec)$, all interwell coupling is destroyed, the quantum states are localized to individual wells and electrons tunnel from one well to another until they reach the collecting electrode. At intermediate fields, the miniband is replaced by a Wannier-Stark ladder and electron transport evolves gradually toward hopping as the coupling between wells diminishes\cite{De95}.

\begin{figure}[thp!]
  \centering
  \includegraphics[width=3.6 in,angle=0,clip=]{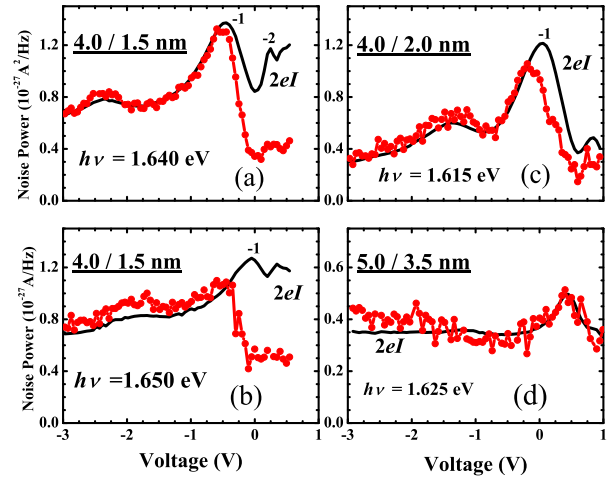}
  \caption[Fano and the current ]{\label{fig:fig3}   Experimental $2eI$ (solid line) and noise (connected circles) of the low-temperature (6K) photocurrent characteristics of various $p-i-n$ GaAs-GaAlAs heterojunctions with a superlattice in their intrinsic regions. The junctions differ only in the well/barrier thickness of their superlattices. The current maxima (denoted by $n = -1, -2$) correspond to resonant absorption involving Stark-ladder states in the conduction band. In (a) and (b), corresponding to a 4.0/1.5 nm superlattice, for voltages larger than -0.5V, the experimental shot noise is much smaller than the Poissonian value ($2eI$), whereas for smaller voltages the two values are practically the same, regardless of the excitation wavelength.  The noise reduction was much less pronounced in a 4.0/2.0 nm structure, (c), and was absent in a 5.0/3.5 nm structure, (d).} 
\end{figure}

The heterostructures from which the photodiodes were made were grown by molecular beam epitaxy using $n^+$GaAs as a substrate. In each heterostructure an undoped region consisting of a superlattice terminated by 600 \AA \, of Ga$_{0.65}$Al$_{0.35}$As on each side was clad by 
heavily doped ($2\times 10^{18}$ cm$^{-3}$) $n^+$ and $p^+$ GaAs electrodes. The superlattice was formed by a large number of periods (twenty to seventy, depending on the sample) of alternating GaAs wells and Ga$_{0.65}$Al$_{0.35}$As barriers, with thickness ranging from 2.0 to 5.0 nm for the wells and from 1.5 to 4.0 nm for the barriers. The extreme values for  $D$ were 5.5 nm and 8.5 nm  and their miniband widths were 98 meV and 15 meV, respectively. The corresponding characteristic fields were 190 and 20 kV/cm, respectively. In total, nine different heterostructures were prepared, from which 400 $\mu$m-diameter photodiodes were fabricated using standard processing techniques. 

Optical excitation was provided by the light from a tungsten lamp passed through a monochromator whose spectral resolution was somewhat sacrificed in order to achieve photocurrent levels sufficient to make noise measurements reliable. A light blocker was placed in front of an optical cryostat where the samples were maintained at a temperature of 6K. Because of the low photocurrent (of the order of nanoamperes) and the high capacitance (about 80 pF) of the diodes, the current was amplified by a special voltage amplifier, whose output was fed to a spectrum analyzer and a lock-in amplifier in parallel. The photocurrent noise was determined differentially, that is, by subtracting the spectral density outputs from the spectrum analyzer when the light blocker was in and out of the path of the light. 

The photocurrent of several superlattices as a function of the total voltage applied to the $p-i-n$ diodes is shown in Fig.2. (For reasons that will soon become clear, in the figure the current is multiplied by $2e$.) The most salient feature of the $I-V$ characteristic is the presence of several maxima and minima, with regions of positive and negative differential conduction between them. The origin of this oscillatory behavior is well known: a maximum appears in the photocurrent whenever there is resonant absorption of light between a localized quantum-well state in the valence band and a delocalized state in the conduction band of the superlattice\cite{Agu89}.
 
For a fixed photon energy, there are resonances at certain electric fields, each resonance involving a certain Wannier-Stark state in the conduction band. If $W$ is the length of the depletion region of the $p-i-n$ heterojunction in which the superlattice is immersed and $h\nu$ is the photon energy, then the resonance voltages are approximately given by $V_n = V_b + W(h\nu - E_0)/eDn$, where $V_b$ is the built-in voltage of the junction, $E_0$ is the optical-transition energy in the very high-field limit, and $n$ is the index of the Wannier-Stark state\cite{Agu89}.

The noise spectral power $S$ exhibits an oscillatory behavior similar to that of the photocurrent, with maxima and minima occurring at the same voltages, as shown in Fig.2. Howewer, a direct comparison between $S$ and $2eI$ reveals that the proportionality constant between the two quantities, in general, is not the same throughout the full voltage range. For instance, for the 4.0/1.5 nm superlattice [Fig.2(a)] the Fano factor is 1 for $V \leq  -0.5$V, but it is significantly smaller for larger voltages, down to $\approx 0.40$ as $V$ approaches $V_b$, independently of the excitation energy [see Fig.2(b)]. The voltages of the oscillation's extrema depend on the excitation energy, but the crossover voltage at which $F$ changes from its minimum to its maximum value does not. The last two frames of Fig.2, (c) and (d), depict the shot noise in comparison with $2eI$ for two superlattices with different well/barrier thicknesses. While for the 4.0/2.0 nm superlattice a noise reduction is still noticeable at positive voltages (although less pronounced than in the 4.0/1.5 nm case), for the 5.0/3.5 nm superlattice the noise is practically Poissonian throughout the entire voltage range.

\begin{figure}[thp!]
  \centering
  \includegraphics[width=3.6 in,angle=0,clip=]{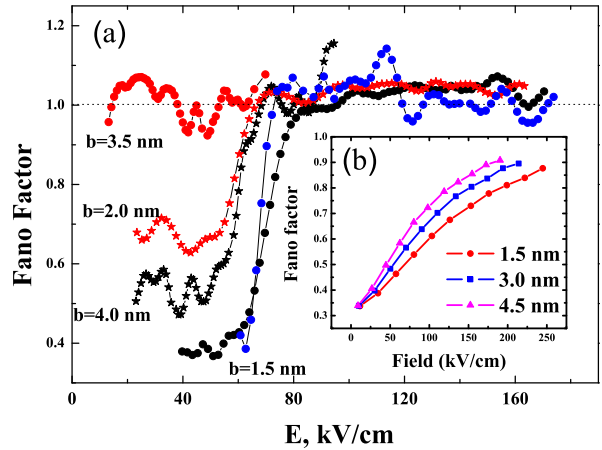}
  \caption[Noise and the current ]{\label{fig:fig3}  Experimental Fano factor as a function of electric field, for five superlattices with the same well width (4.0 nm) and barrier widths ranging from 1.5 to 4.0 nm. (We show data for two superlattices with 1.5 nm barrier width that were grown several months apart from each other.) At moderately high fields ($>$ 80 kV/cm) $F\approx 1$ for all superlattices. At low fields $F$ is the smaller, the narrower the barrier width, with the exception of the 4.0 /4.0 nm heterostructure. A semiclassical multi-barrier model predicts a low-field value of $F = 1/3$, independent of the barrier width, and a gradual increase toward $F = 1$ at very high fields, as shown in (b), where we summarize the calculated dependence of the Fano factor on electric field for three superlattices with the same 4.0 nm well width and different barrier widths: 1.5 nm, 3.0 nm or 4. 0 nm. }
\end{figure}

To establish a meaningful comparison of the behavior of the noise among various superlattices, we have summarized in Fig.3 the dependence of the Fano factor on electric field for superlattices that have the same well width (4.0 nm) but different barrier width, from 1.5 nm to 4.0 nm. The most notable features in the plot are the following. First, for all the superlattices the maximum Fano factor, $F_{max}$, is 1.0, within experimental uncertainty. Second, the value of the minimum Fano factor, $F_{min}$, varies from one superlattice to another, within the range 0.4 to 1.0; the trend is for $F_{min}$ to increase with barrier width. Third, the field region at which the crossover for $F$ occurs is relatively sharp, being slightly sharper the wider the barrier width. 

These three features are quite general to the nine superlattices we have studied, if we use the miniband width, $\Delta$, as the single defining parameter: $F_{max}$ is 1.0 regardless of $\Delta$; $F_{min}$ decreases with increasing $\Delta$; and the crossover between $F_{min}$ and $F_{max}$ occurs at lower field, and is sharper, the smaller $\Delta$. There are a few superlattices that deviate from that behavior, though. The most outstanding exception is the 4.0/4.0 superlattice shown ing Fig.3, whose $F_{min}$ was found to be $0.55 \pm 0.08$. In contrast, for a 5.0/3.5 structure (for which the miniband width $\Delta = 0.015$ eV is comparable to that of the 4.0/4.0 structure) we found $F_{min}\approx 1$. Because the nine superlattices were grown in a span of several years, it is difficult to pinpoint the origin of the deviations from the general trend. At this point, we can only say that superlattice ``quality", evidenced experimentally by several oscillations in the photocurrent\cite{Agu89}, does not seem to be the determining factor for $F_{min}$.

The two limits, $F_{min}$ and $F_{max}$, of the Fano factor can be understood qualitatively by considering the potential profile of a superlattice in the extreme-field regimes. At very high fields, the profile is "tilted" so much that electrons injected optically into a quantum well in the conduction band in practice tunnel through a single barrier (see Fig.1). In this case, the transport process is uncorrelated, and the Fano factor should be one, as we have found experimentally. At low fields, the wells are barely tilted so that electrons tunnel through a large set of almost identical barriers, and the Fano factor should be drastically reduced, down to 1/3 as the number of barriers approaches infinity\cite{De95}.

Beyond this qualitative understanding, the agreement between experiment and theory breaks down. First of all, experimentally the $F\rightarrow 1$ limit is reached at fields quite below the high-field regime. For instance, for the 4.0/1.5 superlattice $F$ is already 1 when the field is about 90 kV/cm (see Fig.3), which is significantly lower than $E_c$ (190 kV/cm) for that superlattice. At that comparatively moderate field, optically pumped electrons effectively "see" several quantum barriers ahead of them as they tunnel through; consequently, the Fano factor should still be significantly smaller than one. In the low-field limit, existing theories predict that $F$ should approach 1/3 as long as the barriers are identical and their number large, regardless of the barrier width\cite{De95}. Experimentally, this seems to be true mostly when the barriers are narrow and, therefore, the interwell coupling strong, but when the barriers are relatively wide $F$ is still close to one at very low fields.

It could be argued that the assumption of identical barriers is not really applicable to our experiments. After all, the field not only reduces the number of barriers that participate in a tunneling event but it also makes them dissimilar as far as tunneling is concerned. For a better comparison with the results summarized in Fig.3, we have adapted an existing one-dimensional calculation [11] to determine the Fano factor of a multi-barrier structure as a function of electric field. We proceeded the following way: for a given field, first we used the transfer-matrix method to determine the tunneling probability of an electron through each of $N (N = 8)$ barriers, individually. At zero field the shape of all the barriers is identical, but since the field displaces them ''vertically'' with respect to each other, a tunneling electron will see each barrier differently, experiencing an increasing tunneling probability as it encounters successive barriers in its motion against the field. Once we had determined the individual probabilities, we used theoretical expressions \cite{De95} to compute $F$.

A summary of our calculations for three different superlattices with 4.0 nm well width is shown in Fig.3(b). As expected, when $N$ is relatively large and all the barriers are identical, at zero field the Fano factor is 1/3, regardless of the superlattice period. As the field increases, F increases monotonically, but the rate of increase is the faster the larger the period (or, equivalently, the smaller $\Delta$). For the superlattice with the longest period, at 200 kV/cm $F$ is 0.9, whereas for that with the shortest period $F$ is 0.8, and approaches saturation very slowly. We note that the critical fields for the superlattices are 20 kV/cm and 190 kV/cm, respectively, so that even when $E$ is ten times $E_c$, the calculated value of $F$ is still less than one. This result is in sharp contrast with our experimental finding (Fig.3(a)) that $F$ is already one even when $E$ is smaller than $E_c$. The other crucial difference, as discussed above, is that, experimentally, at zero field $F\!\!\rightarrow\!{1/3}$ only for the superlattice with the shortest period.

At the moment we do not have an explanation for these discrepancies, which suggest that something is amiss in the simple model we have used when trying to account for the behavior of three-dimensional heterostructures. The measurements described above also show that by substituting a superlattice for the light-absorbing well of a quantum-well photodiode it is in principle possible to reduce significantly the shot noise of the device. We have found that reduction to be the largest at zero field because it is then that the tunneling probability is the same for all barriers. It should be possible to design a multi-barrier structure in which the individual barrier widths and/or heights are arranged in such a way that the optimum noise condition is achieved for an operating field different from zero.

We are grateful to J. M. Hong, who prepared the heterostructures whose noise properties we have reported here. This work has been sponsored by the National Science Foundation under Grant No. DMR-0305384.

\end{document}